\documentclass[aps,prb,twocolumn,superscriptaddress]{revtex4-1}

\usepackage{hyperref}
\usepackage{graphicx}
\usepackage{topcapt}
\usepackage{booktabs}
\usepackage{longtable} 
\usepackage{verbatim}
\usepackage{lineno}

\newcommand{\baxis}{$b$-axis}
\newcommand{\aaxis}{$a$-axis}

\newcommand{\BFA}{BaFe$_2$As$_2$}
\newcommand{\ub}{$\mu_B$}
\newcommand{\iA}{\AA$^{-1}$}
\newcommand{\SQ}{S(Q)}
\newcommand{\Q}{Q}
\newcommand{\Orth}{$\mathcal{O}$}
\newcommand{\GSAS}{{\sc GSAS}}
\newcommand{\TN}{T$_N$}
\begin{document}

\title{Local structural variation as source of magnetic moment reduction in \BFA}


\author{Jennifer L. Niedziela}
\email[]{jniedzie@utk.edu}

\affiliation{Department of Physics and Astronomy, University of Tennessee - Knoxville, Knoxville, TN 37996}
\affiliation{Oak Ridge National Laboratory, Oak Ridge, Tennessee, 37831}

\author{M. A. McGuire}
\affiliation{Oak Ridge National Laboratory, Oak Ridge, Tennessee, 37831}

\author{T. Egami}
\affiliation{Department of Physics and Astronomy, University of Tennessee - Knoxville, Knoxville, TN, 37996}
\affiliation{Oak Ridge National Laboratory, Oak Ridge, Tennessee, 37831}
\affiliation{Department of Materials Science and Engineering, University of Tennessee - Knoxville, Knoxville, Tennessee, 37996}


\date{\today}

\begin{abstract}
We report time-of-flight neutron powder diffraction results on stoichiometric \BFA.  Pair distribution function analysis shows that the orthorhombic distortion in the $a-b$ plane at short distances are significantly different from the average lattice distortion, indicating local variations in the lattice at the short-range.  We propose that this local variation reflects a high density of nano-twins, short-ranged structures which locally affect the magnetic alignment. This results suggests that the discrepancies between the observed and calculated magnetic moments in \BFA\ arise partly from short-ranged variation of the lattice in the $a-b$ plane.   
\end{abstract}


\maketitle


\section{Introduction}
Since the discovery four years ago of high-temperature superconductivity in iron based compounds (FeSC)\cite{Hosono:2008}, much progress has been made to identify the relevant physics of these materials. Nevertheless, many questions about the superconducting and magnetic properties remain, many of these dealing directly with the questions of the influence of structural transitions to the electronic properties of the system.  We have used pair distribution function analysis to describe the local atomic structure of \BFA, a parent compound to several families of FeSC.  In this Letter, we report the presence of local atomic disorder in \BFA, and discuss the impact of the disorder on the magnetic properties.  By comparing magnetic properties measured with spectroscopic probes, calculations, and neutron diffraction, we underscore the importance of the structural contribution to the electronic properties of the system. 

\BFA\ is a tetragonal paramagnet above 140 K at ambient pressure. Below T $\approx$ 140~K, the symmetry of \BFA\ is lowered to an antiferromagnetic (AF) orthorhombic state [Fig. \ref{fig:Rietveld30K}a]. This distortion changes the planar layer of iron atoms from squares to rectangles by expanding along the \aaxis\ and contracting along the \baxis\ by 0.01 \AA \cite{Rotter:SDW, Mandrus:BFA}. Superconductivity is induced in \BFA\ through suppression of the magnetic ground state via a number of mechanism (See Ref. \cite{Mandrus:BFA} for review), and the magnetic and superconducting behaviors of FeSC have clear dependence on the structural properties \cite{Lee:CriticalAngle,Mizuguchi:CriticalSeparation,Egami:SpinLattice}.  In the local moment picture, this small orthorhombic distortion induces large anisotropy in the magnetic exchange constants as determined from spin wave fittings, with AF $J_{1a}$ and FM $J_{1b}$ having an opposite sign [Fig. \ref{fig:Rietveld30K}b] \cite{Harriger:2011PRB, Han:Pnictides}.    Pair distribution function (PDF) analysis of \BFA\ and related compounds by x-rays shows no evidence local distortion in the K-doped regime\cite{Joseph:KBFA}, but neutron analysis shows the presence of spin-ladder structures in chalcogenides\cite{Caron:SpinLadder}, and an eventual ambiguity in the structural assignment of the Fe(Se,Te) class of materials\cite{Louca:FeTeSe}.

Density functional theory (DFT) calculations predict values for the magnetic moment from 1.75 to 2.4~\ub/Fe, \cite{Singh:BFA, Boeri:MagnetismBFA,Han:Pnictides} and show that the local Hund's coupling is primarily responsible for large magnetic moment. \cite{Han:Pnictides}  The value of the local moment observed from core electron spectroscopy is 2.1 \ub\ for SrFe$_2$As$_2$. \cite{Mannella:LSMPRB}  On the other hand, the magnetic moment of \BFA\ determined from neutron diffraction measurements range from 0.87 to 1.04~\ub/Fe, \cite{Huang:BFAOrder, Wilson:BFA} and sum-rule integrations over the spin excitation spectrum yield 1.04  \ub. [\onlinecite{Menshu:Nature}]
$\mu$SR and M\"{o}ssbauer studies of \BFA\ show an ordered iron moment of 0.4\ub.[\onlinecite{Aczel:PRB, Rotter:SDW}]

Part of the discrepancy between the core electron spectroscopy measurement and other spectroscopic techniques of nearly a factor of 2 could be dynamic, because the time scale of spectroscopic measurement is very short, and the DFT calculations do not include quantum fluctuations.  However, it is possible that the discrepancy originates also from the fact that the spectroscopic measurement measures local moment, whereas the diffraction measures the average moment.  This is the point we investigate in this work.    
   
We have conducted a study of the \BFA\ system using neutron powder diffraction and pair distribution function (PDF) analysis. Information about local distortion and local atomic correlation is present in the high \Q\ diffraction data, for instance as diffuse scattering. The PDF is the Fourier transform of \SQ, and retains all of the information regarding local structural distortions and atomic correlations, allowing one to incorporate this information into a comprehensive model of the local structure of a system \cite{EgamiBook,Goodwin:PDFReview}.

\section{Experiment}
\begin{figure}
\includegraphics[width=3.4in]{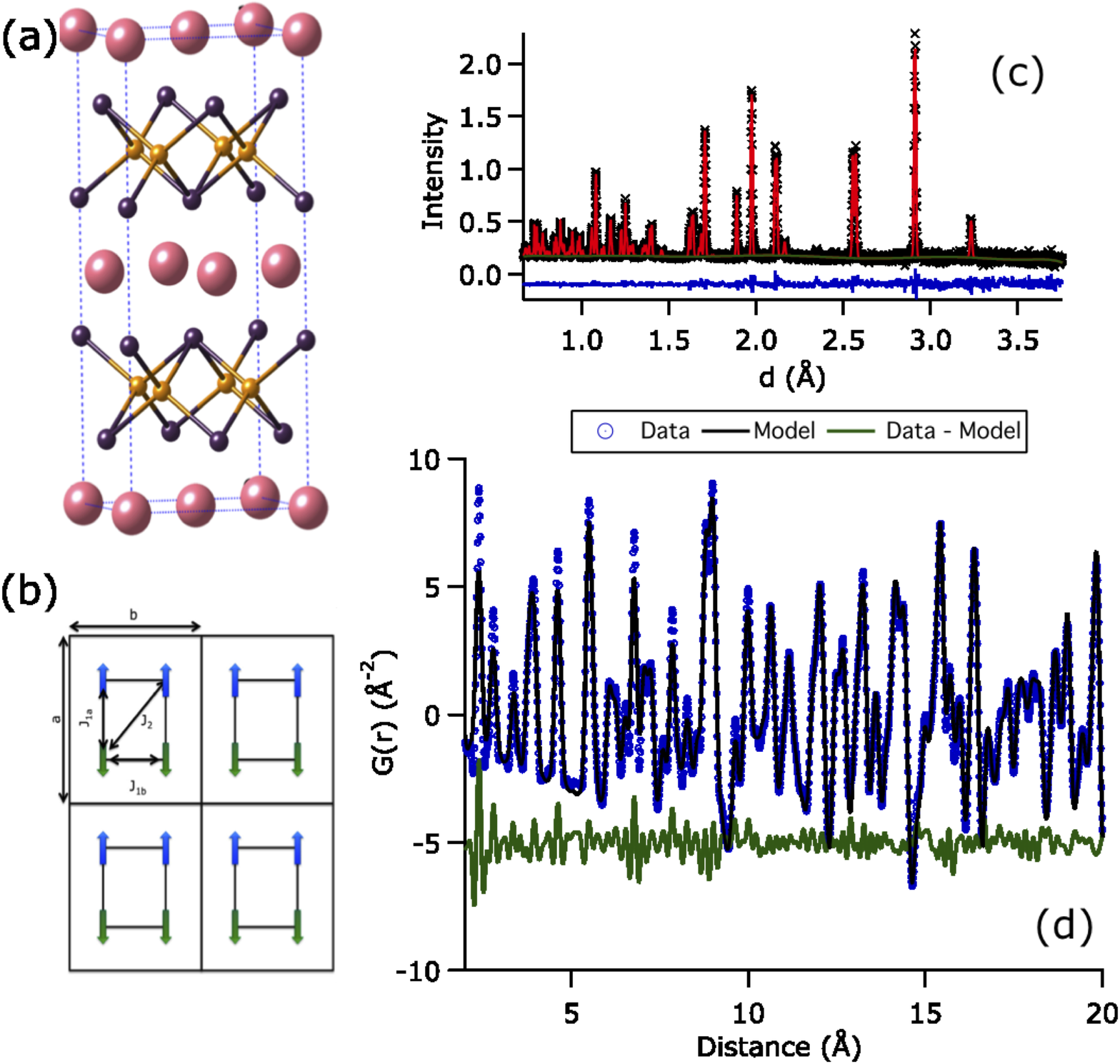}

\caption{(color online) (a) Orthorhombic crystal structure and spin alignment of \BFA\ as suggested by neutron scattering experiments. (b) Magnitude of experimentally determined exchange constants decreases with the relationship $J_{1a}$ $>$ $J_2$ $>$ $J_{1b}$. (c) Rietveld refinement pattern for \BFA\ at 30K.  (d) PDF data compared to the  Rietveld structural model.  The PDF for the Rietveld obtained structural model deviates most substantially at the shortest range, exhibiting a clear crossover at 10 \AA.    Refinement parameters are summarized in Table \ref{tab:BFAparams}. }
\label{fig:Rietveld30K}
\end{figure}

Neutron time-of-flight powder diffraction experiments were conducted at the Manuel Lujan Neutron Scattering Center at Los Alamos National Laboratory using the NPDF powder diffractometer \cite{Proffen:NPDF}.  A sample of \BFA\ was prepared using powder synthesis methods (\TN\ ~=~138~K). Samples were sealed in a vanadium sample can in a helium glovebox, and placed into a cryostat for temperature controlled studies.  The diffraction data were analyzed by Rietveld refinement as implemented in \GSAS \cite{GSAS},  and the results from the Rietveld refinement are shown in Fig. \ref{fig:Rietveld30K}c.  Refinement parameters for Rietveld analysis included lattice parameters, absorption, isotropic atomic displacement parameters (ADP), and the $z$ position for the As atoms. The background was modeled using a Chebyshev polynomial.

\begin{figure}
\includegraphics[width=3in]{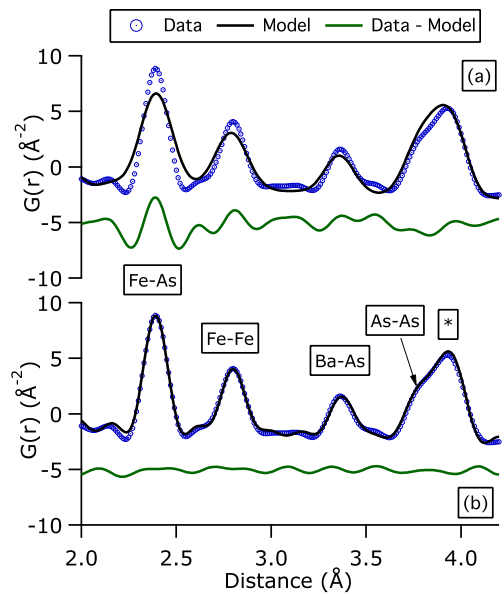}
\caption{\label{fig:SRPDFRietComp}color online) Comparison of short-range (2-4.2 \AA) PDF data to the (a) Rietveld and (b) refined PDF structural models with interatomic distances labeled. The Rietveld structural model deviates in peak positions at the Fe-As, Fe-Fe, and As-As distances, while the peak heights indicating the coordination in the local environment are not accurate.  In b) the lattice parameters and isotropic thermal parameters are free parameters to the fit, improving the fit markedly.  The * in (b) marks the distance of the tetragonal $a$ lattice parameter.}
\end{figure}

\begin{figure}
\includegraphics[width=2.5in]{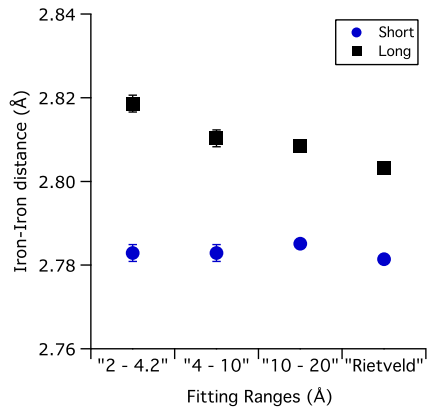}
\caption{(color online) Iron bond lengths from fit PDF for \BFA\ at 30K.  The length of the long side of the plaquette gradually decreases as a function of fitting range reflecting the degree of short-range distortion in \BFA. }
\label{fig:FeFeBondsBFA}
\end{figure}

PDF data were produced using standard methods \cite{PDFGetN} and quantitative analysis was done with PDFgui \cite{PDFGui}.  Fourier transforms were terminated at 35 \iA.  The Rietveld structural model was the starting model for the PDF analysis.  The measured PDF was found to deviate substantially from the PDF calculated for the model obtained by the Rietveld refinement [Fig. \ref{fig:Rietveld30K}d].  Agreement is poorest for the short-range (2-4.2 \AA) [Fig.~\ref{fig:SRPDFRietComp}a] and improved at long-range (10-20 \AA)  [Fig. \ref{fig:Rietveld30K}d].  Allowing lattice parameters and isotropic ADPs to be free parameters to the fit, we find that the fit in the short-range is markedly improved [Fig. \ref{fig:SRPDFRietComp}b] and that different lattice parameter values describe the system for the different fitting ranges.  Consequently the Fe-Fe distance changes as a function of fitting range [Fig. \ref{fig:FeFeBondsBFA}]. At the same time the isotropic ADPs of Fe and As [Table \ref{tab:BFAparams}] exhibit a monotonic increase in value as a function of fitting range.

%
%
We characterize the behavior of the Fe-Fe distance in the \textit{a-b} plane, $R_{\text{Fe-Fe, a}}$ and $R_{\text{Fe-Fe, b}}$ through a quantity orthorhombicity, defined as: $\mathcal{O}=\frac{ R_{\text{Fe-Fe, a}}- R_{\text{Fe-Fe, b}}}{ R_{\text{Fe-Fe, a}}}$.  The value of the \Orth\ is maximal at the shortest fitting range at 1.38\%.  For comparison, the value determined from the Rietveld refinement is 0.78\%, representing a 60\% increase in \Orth\ at the short-range.  

%
%

%
%

%
%

 \begin{table*}
 \begin{ruledtabular}

   \centering
   \topcaption{PDF and Rietveld refinement results for \BFA\ at 30K; $u$ values are isotropic atomic displacement parameters, calculated from the trace of the atomic displacement tensor.} 
   \begin{tabular}{@{} c c c c c c c c  @{}} 
      \toprule
Range (\AA)    & a (\AA)   & b(\AA)    & c(\AA)& u-Ba (\AA$^2$) &     u-Fe (\AA$^2$) &    u-As (\AA$^2$)  \\
\hline
2.0-4.2     &5.63(2) &5.56(2)    &12.925(4) &    0.0026(3)&    0.0017(3)    &0.0008(3)    \\
4.0-10.0 &    5.620(5)&    5.566(5)    &12.953(6)    &0.0017(6)&    0.0022(2)    &0.0015(3)    \\
10.0-20.0    &5.617(1)&    5.570(1)    &12.951(3)    &0.0032(6)&    0.0034(2)&    0.0016(3)   \\
\midrule
Rietveld &  5.6065(2)  & 5.5627(7) & 12.9284(4) & 0.0025(5) &    0.00343(29) & 0.00330(32)  \\    
      \bottomrule
   \end{tabular}
  
   \label{tab:BFAparams}

\end{ruledtabular}

\end{table*}

\section{Discussion}
\begin{figure}
\includegraphics[width=3.38in]{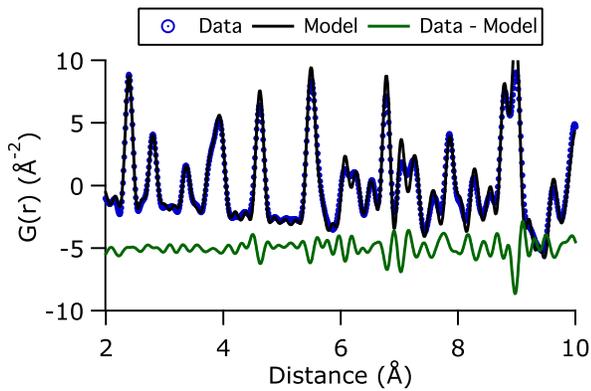}
\caption{\label{fig:SRFailure}(color online) Application of the short-range structure applied to the medium- (4-10 \AA) range structure.  The short-range structural model fails to describe the medium-range structure, indicating that the short-range structure of \BFA\ is distinct.}
\end{figure}

The poor fit of the Rietveld derived PDF and resultant increase in \Orth\ at the short-range shows that the local structure of \BFA\ seen by PDF is different from the average structure determined by the Rietveld analysis.  The structural heterogeneity is also seen in the monotonically increasing isotropic atomic displacement factors.  Apparently the iron plaquette is locally more distorted than the average structure [Fig. \ref{fig:FeFeBondsBFA}).  Conversely, the short-range structure which fits the PDF at short distances does not describe the medium- or long-range structure [Fig. \ref{fig:SRFailure}], showing that the local structure is distinct from both the medium- and long-range structure.  

We note that the previous X-ray study (Ref. \onlinecite{Joseph:KBFA}) does not report local distortions in \BFA, preferring to treat the system with the inclusion of anisotropic thermal factors. We believe the discrepancies between these reports arises from the difference in experimental conditions: their \Q\ range is limited to 25 \iA\  \onlinecite{Joseph:KBFA}, whereas our neutron measurements reach a higher \Q\ range, up to 35 \iA\ providing better real-space resolution.  

\begin{figure}[!t]
\includegraphics[width=3.3in]{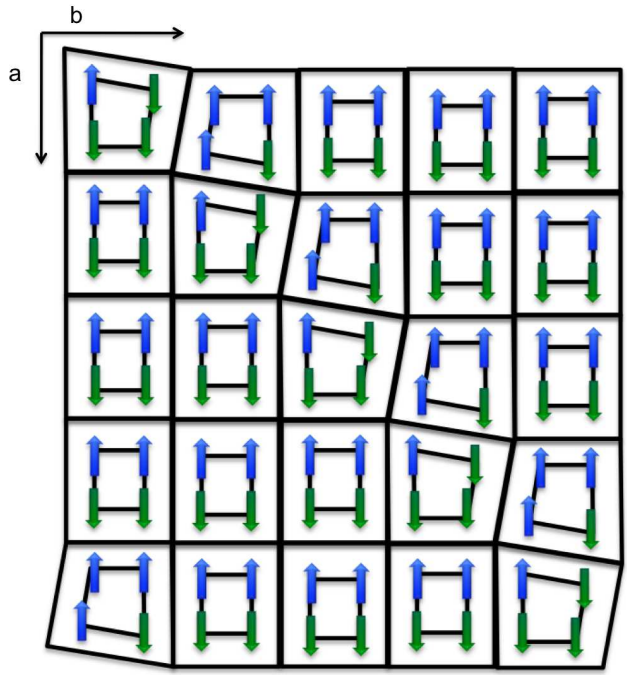}
\caption{\label{fig:Frustrated}(color online)  Suggested nano-twins of \BFA\ with corresponding alteration of magnetic alignment. Nano-twins locally reorient the iron plaquette, forcing the spins to realign into nano-scale AF domains.  This results in an overall reduction of the magnetic moment of the system as measured by neutron diffraction. }
\end{figure}

We explain the difference between the local and global Fe-Fe distances using a nanoscale twinning model. In this model the local structure has orthorhombic distortion with \Orth\ = 1.38\%. If the volumes of the nano-twins in two orientations are equal, the macroscopic distortion will be zero. In our model the volumes of the twins are not equal, resulting in a reduced, but non-zero, strain of \Orth\ = 0.78\%.  An example of such nano-twins is shown in Fig. \ref{fig:Frustrated}.  Regular structural modulation as shown in Fig. \ref{fig:Frustrated} will produce superlattice diffraction. Absence of superlattice diffraction with \BFA\ suggests that the modulation is not regular, but is random, and we note that random modulation will not result in broadening of diffraction peaks \cite{Teller:Disorder}. 

The presence of such nano-twins should have a profound effect on the magnetic structure, because the magnetic exchange parameters are positive (AF) along the direction of longer Fe-Fe distance ($a$ axis), and negative (FM) along the direction of shorter Fe-Fe separation ($b$ axis). \cite{Harriger:2011PRB,Han:Pnictides}  Therefore the direction of the AF spin alignment should largely follow the orientation of the twins. The simplified model shown in Fig. \ref{fig:Frustrated} shows that the realignment of the spins along the nano-twin boundaries results in nano-scale AF domains and weakening of the magnetic diffraction peaks observed by neutron diffraction.  If, as is shown in the figure, the nano-twin with the volume fraction of 20\% reduces the average orthorhombic distortion by 40\%, the AF sublattice moment should also decrease by 40\%.  Again, the model shown in Fig. \ref{fig:Frustrated} is overly simplified and the AF domain walls may have some width, resulting in smoother local variations in the spin orientation. However, it is obvious that the complex behaviors of the nano-twins could easily result in a significant moment reduction.

We note also that the value of the moment obtained from M\"{o}ssbauer\cite{Rotter:SDW} is consistent with the value obtained by $\mu$SR\cite{Aczel:PRB}, a value of 0.4 \ub, which introduces a discrepancy of a factor 2 between the slower timescale methods and the value of the moment as measured by neutron diffraction.  We can infer from the reported information that the expected error on these values would still place the value of the moment less than that reported by neutron diffraction.  Further, $\mu$SR studies report the presence of inequivalent magnetic sites in the unit cell of \BFA\cite{Aczel:PRB}, raising a question of the true distribution of the spin density in the system, and emphasizing the important dynamic aspects of the magnetism in this system.

The nano-twin model is consistent with results in several published works. Above \TN, the system exhibits broad quasi-elastic signals at the AF wave vector in inelastic neutron studies \cite{Harriger:2011PRB,Matan:BFASpinExcitation}, indicating retention of strong in-plane medium-range spin order in the paramagnetic state in spite of a loss of inter-planar correlation. As the temperature of the system is lowered toward \TN, the spatial range of spin correlation increases
and the fluctuation frequency is lowered. Thus spin correlation couples with the
lattice, resulting in local distortion. This effect is represented by the softening of the transverse phonon mode near \TN. \cite{NiedzielaBFA} The softening occurs over a finite range of inelastic momentum transfers $\mathbf{q}$, so the correlation length of the fluctuations never diverges as in the standard second-order phase transition. Indeed this transition has first-order signature \cite{Kofu:BFA,Matan:BFASpinExcitation,Rotundu:PhaseTransition}, suggesting the local nature of the transition.  Further, the resulting increase in the magneto-structural transition temperature by compressing the orthorhombic \baxis \cite{Dhital:BFAStrain} is naturally explained by this model.  Applying uniaxial strain to the \baxis\ induces partial de-twinning, resulting in the increase in the volume fraction of the twin with one sense and the intensity of the magnetic diffraction peak. 

Additionally, systematic differences in the magnetic moment from neutron diffraction are observed in \BFA\ as a function of cooling temperature.  As observed in Ref. \onlinecite{Lumsden:Review}, the level of hysteresis in the magnetic transition is highly sample dependent, in addition to depending on the rate of cooling used in the experiment\cite{Huang:BFAOrder, Wilson:BFA,Kofu:BFA}, highlighting the instability of the formation of the magnetic structure in these systems.	

%

%

\section{Conclusion}

To conclude, we present evidence for a local structural variation of the \BFA\ system, in the form of nano-twins.  The presence of the nano-twins introduces nano-scale AF domains in magnetic order in the \BFA\ system, lowering the average static magnetic moment of the system seen by neutron diffraction. Observations are made in the as-grown compound \BFA, absent any distortion that might be expected from chemical substitution or alloying, showing that the system exhibits a fundamental structural and magnetic frustration.  By focusing on the parent compound we decoupled the effect of alloying and substitutional disorder from the interpretation.  But there is ubiquitous evidence of similar local distortions across many FeSC compounds and this will be discussed elsewhere.  This result addresses an outstanding question in the study of high-temperature iron-based superconductors, explaining in part that the source of the discrepancy between moments measured by the spectroscopic probes, DFT calculations, and neutron diffraction is intrinsic structural distortion in the pnictide compounds.

\section{Acknowledgements}
The authors would like to thank Pengcheng Dai and Stephen Wilson for useful discussions. This work was supported by the Office of Basic Energy Sciences, US Department of Energy through the EPSCoR Grant, DE-FG02-08ER46528 (JN and TE), and the Basic Energy Sciences, Materials Sciences and Engineering Division (MAM). Work at the Manuel Lujan Neutron Science Center at Los Alamos National Laboratory was supported by the U.S. Department of Energy, Office of Science, Office of Basic Energy Sciences.




\end{document}